# VARIABLE THRESHOLD MOSFET APPROACH(THROUGH DYNAMIC THRESHOLD MOSFET) FOR UNIVERSAL LOGIC GATES

K. Ragini,[1] Dr. M. Satyam,[2] and Dr. B.C. Jinaga,[3]

[1] G. Narayanamma Institute of Technology & Science, Department of Electronics and Communication Engineering, Hyderabad, India

ragini_kanchimi@yahoo.co.in

[2] International Institute of Information Technology, Hyderabad, India

satyam@iiit.ac.in

[3] School of Information Technology, Jawaharlal Nehru Technology University, Hyderabad, India

jinagabc@gmail.com

*ABSTRACT*

*In this article, we proposed a Variable threshold MOSFET(VTMOS)approach which is realized from Dynamic Threshold MOSFET(DTMOS), suitable for sub-threshold digital circuit operation .Basically the principle of sub- threshold logics is operating MOSFET in sub-threshold region and using the leakage current in that region for switching action, there by drastically decreasing power .To reduce the power consumption of sub-threshold circuits further, a novel body biasing technique termed VTMOS is introduced .VTMOS approach is realized from DTMOS approach. Dynamic threshold MOS (DTMOS) circuits provide low leakage and high current drive, compared to CMOS circuits, operated at lower voltages.*

*The VTMOS is based on operating the MOS devices with an appropriate substrate bias which varies with gate voltage, by connecting a positive bias voltage between gate and substrate for NMOS and negative bias voltage between gate and substrate for PMOS. With VTMOS, there is a considerable reduction in operating current and power dissipation, while the remaining characteristics are almost the same as those of DTMOS. Results of our investigations show that VTMOS circuits improves the power up to 50% when compared to CMOS and DTMOS circuits, in sub- threshold region..*

*The performance analysis and comparison of VTMOS, DTMOS and CMOS is made and test results of Power dissipation, Propagation delay and Power delay product are presented to justify the superiority of VTMOS logic over conventional sub-threshold logics using Hspice Tool. . The dependency of these parameters on frequency of operation has also been investigated.*

*KEYWORDS*

*Sub- threshold, Dynamic threshold MOS Inverter, Propagation delay, Noise-margin, Variable threshold MOS Inverter, Power dissipation.*

## 1.0 INTRODUCTION

There have been continual research efforts in recent years to explore the ultra-low power region of operation. These efforts have been effective in applications where ultra low power consumption is a primary requirement and speed is of secondary consideration .The proposed method to achieve ultra low power is to operate the transistor's in sub-threshold region[1,2].Research in sub-threshold region of operation highlights the advantage of operating transistor's below their threshold voltage. This advantage of ultra low power is especially attractive for deployed medical devices(Hearing aids, pacemakers etc),sensors and devices with low processing needs[3].Sub-threshold current has been viewed as a continually increasing source of wasted power in circuits. Instead of wasting the power, the leakage currents can be used as computational current in circuits. This is the basis for sub-threshold logic circuits[4] .Some applications demand a low energy approach, others medium speed and medium power dissipation  Many researchers focused on achieving low power with medium speed.[5,6] Assaderaghi.F[7] I..Chung[8] , A.Drake[9] and M.R.Casu [10] has proposed the circuits, Fig 1(a),1(b) and 1(c) to improve the current drive and showed that the delay has come down compared to CMOS inverters. This in turn has resulted in some increase in the power dissipation

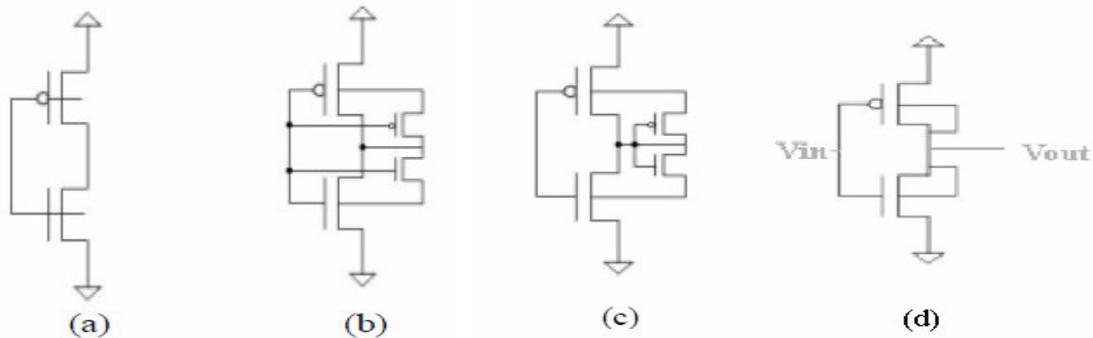

Fig 1(a) Standard DT-CMOS
Fig 1(b) DT-MOS  with augmenting devices
Fig 1(c)  Another topology of DT- CMOS   with   augmenting devices
Fig 1(d) DT-CMOS with Drains   connected to substrate

The additional transistor's  lead to more complexity in the  case of circuits 1(b) and 1(c). Hence Soleimani[11] proposed a new inverter  scheme, 1(d) with drains connected to substrate. He claims that he achieved better power delay product compared to 1(b) and 1(c). In view of this, it is thought that operating the DTMOS architecture , with a proper bias voltage applied between gate and substrate (VTMOS), might result in lowering of operating currents and power dissipation .With a view to examine this conjuncture, circuits based on VTMOS have been conceived and their performance has been analyzed .The aim of this paper is to communicate the results obtained through the simulations on VTMOS  circuits .In the first instance dc transfer characteristics of the MOS devices under VTMOS operation have been obtained through simulations. It has been found that VTMOS circuits do result in lower power dissipation compared to CMOS and DTMOS circuits, while other performance characteristics remain almost the same. Later using these characteristics, VTMOS inverter has been constructed and its performance has been measured through simulation. The analysis is extended to Universal logic gates ( Two-input NAND and Two-input NOR logic gates). Preliminary Investigations carried out on all inverter circuits have been indicated .It has been concluded that the VTMOS can provide the benefits of DTMOS and also results in lower power dissipation with marginal increase in propagation delay.

## 2.0 STRUCTURE OF CMOS, DTMOS AND VTMOS CONFIGURATION

Typical schematic structures of CMOS, DTMOS and VTMOS are given in Fig 2(a), 2(b) and 2(c). In conventional NMOS circuit, Fig. 2(a), the substrate is normally connected to ground or lowest potential in the circuit and in PMOS circuits, the substrate is connected to supply voltage or the highest potential in the circuit. In DTMOS, Fig 2 (b), the substrate is always kept at gate potential. Also, the voltage of each transistor substrate is dynamically adjusted depending on the gate voltage, causing the threshold voltage of the device to adjust dynamically [12,13]. DTMOS devices are efficient because they function as dual threshold logic gates. When a DTMOS transistor is ON, its threshold is lowered increasing the current and decreasing propagation delay. Likewise when the transistor is OFF, the threshold is raised, reducing leakage current and minimizing power and energy dissipation. DTMOS is an excellent scheme to provide ultra-low power with increased speed compared to traditional body biasing in the sub-threshold region [14,15].VTMOS is nothing but an extension of DTMOS in the sense that the substrate voltage differs always by a constant voltage from the gate voltage. As shown in Fig 2(c),by connecting positive bias between gate and substrate for NMOS and negative bias between gate and substrate for PMOS , there is a good reduction of power dissipation in sub-threshold when compared to DTMOS and traditional CMOS. The circuit is named as VTMOS because ,we have used the same DTMOS with a biased voltage between gate and substrate .The voltage of each transistor is dynamically adjusted depending on gate voltage ,causing the threshold voltage of device to adjust dynamically.

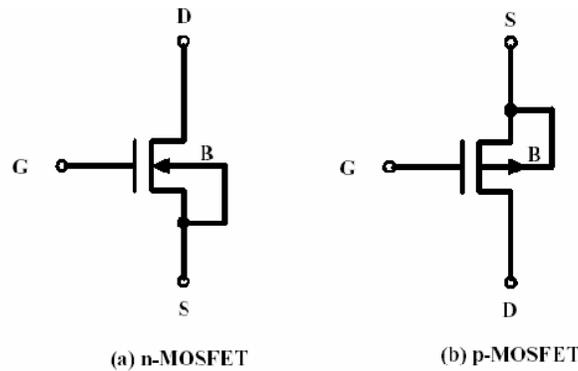

Fig 2(a) - CMOS Structure

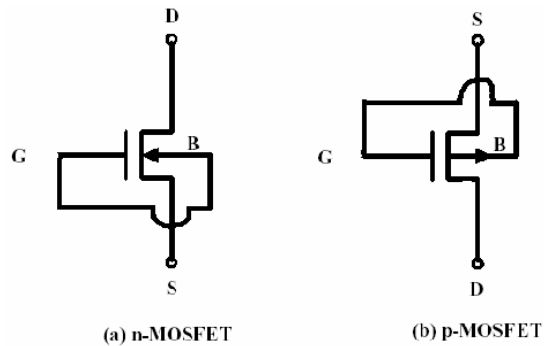

Fig 2(b) - DTMOS Structure

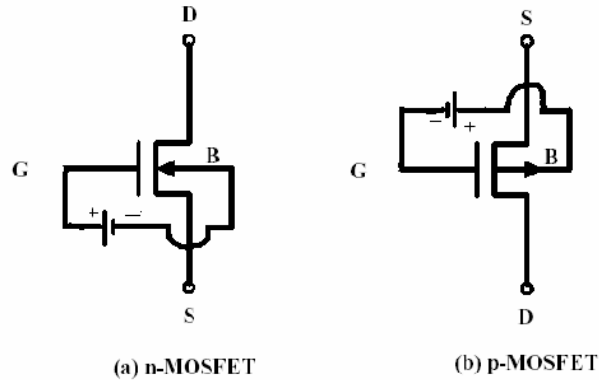

Fig 2(c) - VTMOS Structure

## 2. 1   I-V  characteristics of MOS devices

To evaluate the behavior of NMOS devices under VTMOS operating conditions, the I-V characteristics are measured and are given in  Fig 3(a)and Fig 3(b).It may be observed from  Fig 3(a) and Fig 3(b), that general current levels ($I_{on}$ and $I_{off}$) get reduced with increase in bias voltage, when NMOS gate is positively biased with respect to substrate and in the PMOS case, when gate is negatively biased with respect to substrate .In order to examine the effect of substrate bias on I-V output characteristics  of VTMOS, drain current $I_{ds}$ for different substrate bias voltages,  Keeping gate voltage   $V_{gs} = 0.2v$ have been measured and are given in Fig 3(b).It may be seen that the variation in $I_{ds}$ with drain voltage ,$V_{ds}$ becomes less as $V_{AN}$ is made positive (deep sub- threshold region).  The characteristics may become flat, indicating that the output resistance becomes high. Thus the drain current is less sensitive to variations in drain voltages, which is a welcome feature for application of device in circuits.

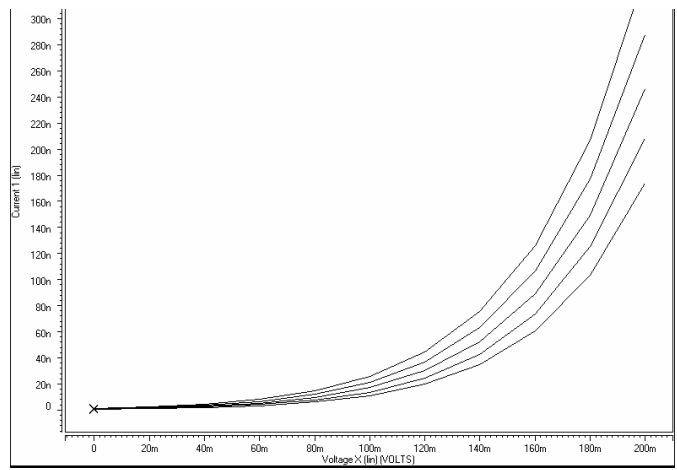

Fig 3(a) -$I_{ds}$ - $V_{gs}$ Characteristics of VTMOS $V_{AN}$ Varying from 0(top) to 0.2 V

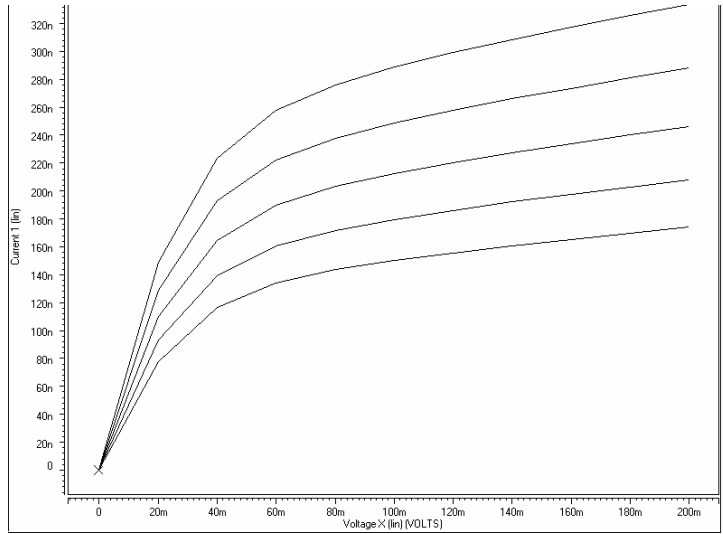

Fig 3(b)- $I_{ds}$-$V_{ds}$ Curves For VTNMOS-$V_{AN}$ Varying From 0(Top) TO 0.2V(Bottom)

## 3.0 VTMOS LOGIC GATES

The transistors for VTMOS logic are chosen from the 65 nm technology [16,17]. The threshold voltage for these devices is 0.22V for VT- NMOS and-0.22V for VT-PMOS. The width of VT-NMOS ($W_N$) and VT-PMOS ($W_P$) is chosen as 200 nm and 400 nm respectively. The supply voltage is taken as 0.2V which is below the threshold of both the devices. For different values of $V_{AN}$ starting from 0 to 0.2V, and corresponding $V_{AP}$ from 0 to - 0.2V, the performance of the VTMOS Inverter, VTMOS Two-input NAND gate and VTMOS Two-input NOR gate- power dissipation, propagation delay, power delay product and frequency response have been obtained, through simulation[18]. When the bias voltage is increased beyond supply voltage, the logic levels are affected. Hence there is a limitation for bias voltage and it should be always below supply voltage In the first instance, the VTMOS Inverter has been constructed as shown in Fig (4) It consists of a PMOS and NMOSFETS, connected in series. The substrate of PMOS is connected to gate through $V_{AP}$, which bias the gate negative with respect to substrate. The substrate of NMOS is connected to gate through $V_{AN}$, which bias the gate positive with respect to substrate. This ensures that both transistors work in the low current region. The voltage transfer characteristics of CMOS and VTMOS Inverter with $V_{AN}$ varying from 0 to 0.2V and $V_{AP}$ varying from 0 to -0.2V have been obtained and is shown in Fig (5). In this case the input is taken in the form of pulse wave varying from 0 to 0.2 V, with a rise and fall time of 25 ns. The frequency of pulse wave is 100Khz..

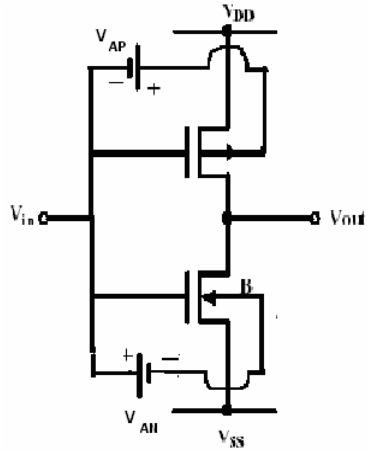

Fig 4 VTMOS Inverter

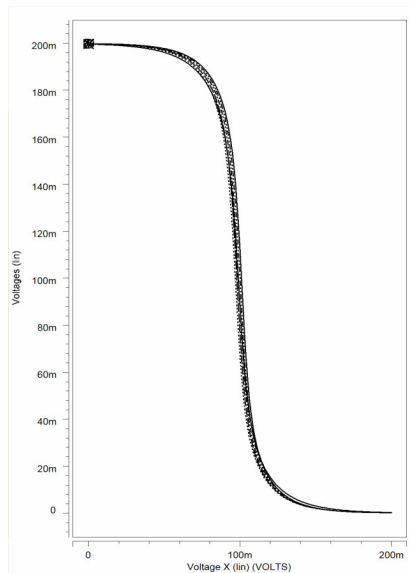

Fig 5 Voltage transfer characteristics of CMOS and VTMOS Inverter ($V_{AN}$ varying from 0 to 0.2v) at 100Khz frequency

The universal logic gates like Two-input NAND and Two-input NOR have been built with the same device parameters .Spice simulated output waveforms of Two-input NAND and Two-input NOR gates are given in Fig 6(a) and Fig 6(b) at a frequency of 400Khz .

38

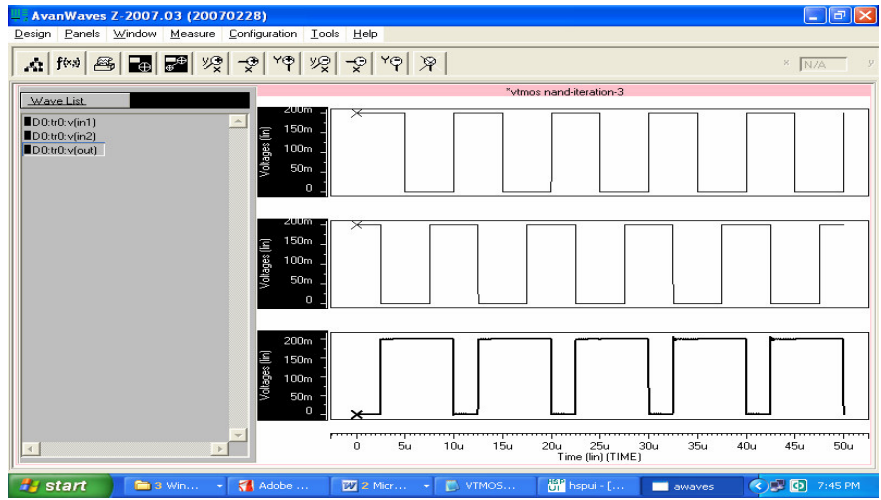

Fig 6 (a) Simulated output waveform of VTMOS Two-input NAND gate for $V_{AN}$ = 0.2V at 400Khz frequency.

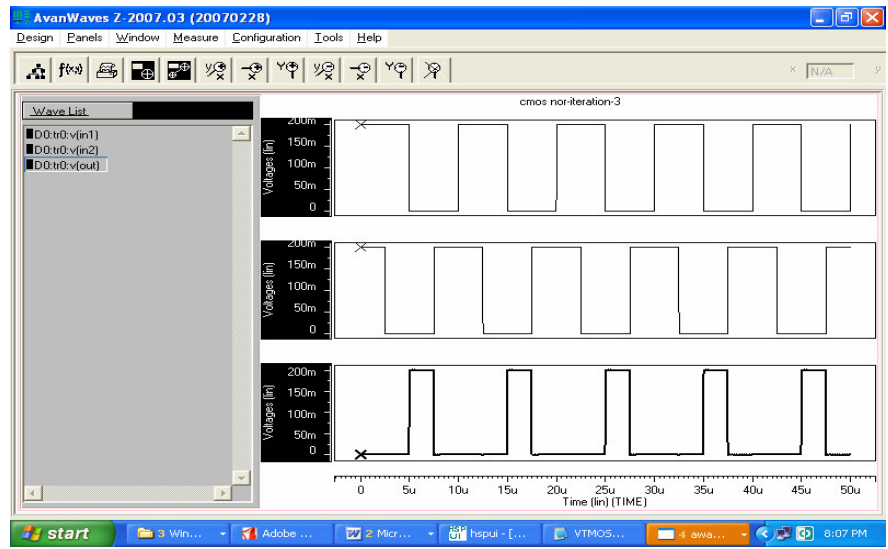

Fig 6 (b) Simulated output waveform of VTMOS Two-input NOR gate for $V_{AN}$ = 0.2V at 400Khz frequency

## 4.0 RESULTS AND ANALYSIS

### 4.1 PERFORMANCE CHARACTERISTICS OF VT MOS CIRCUITS

The variation of propagation delay of VTMOS circuits ( Inverter , Two-input NAND, Two-input NOR) with $V_{AN}$ are compared with CMOS at a frequency of 100 Khz and is shown in Fig 7(a). It may be seen from Fig 7(a), that highest propagation delay in VTMOS circuits is almost equal to propagation delay of CMOS, but at $V_{AN}$ =0, the delay in VTMOS is the lowest and gradually increases to that of CMOS at $V_{AN}$ =0.2V.The average power consumed by the VTMOS Circuits (Inverter ,Two-input NAND ,Two-input NOR) with varying $V_{AN}$ are compared with CMOS at a frequency of 100 Khz and is given in Fig 7(b). From the Fig 7(b), it is observed that there is a slight increase in power dissipation for VTMOS at $V_{AN}=0$, when compared to CMOS. From $V_{AN}=0$ there is considerable reduction in power as $V_{AN}$ is increased. Overall 54% reduction in power can be achieved at frequency=100 Khz in VTMOS circuits for $V_{AN}=0.2V$ compared to CMOS circuits. From Fig 7 (a) and Fig 7 (b), it may be seen that, while the power dissipation decreases with increase in $V_{AN}$, the delay has been found to increase. In order to get the merit of operating the inverter in sub-threshold region, the power delay product [19] is computed and given in the Fig 7(c). It may be seen that, the PDP reduces as we go from CMOS to VTMOS (0.2V). Thus the VTMOS appear to have an edge over the other configuration from the point of power and delay at 100 Khz frequency.

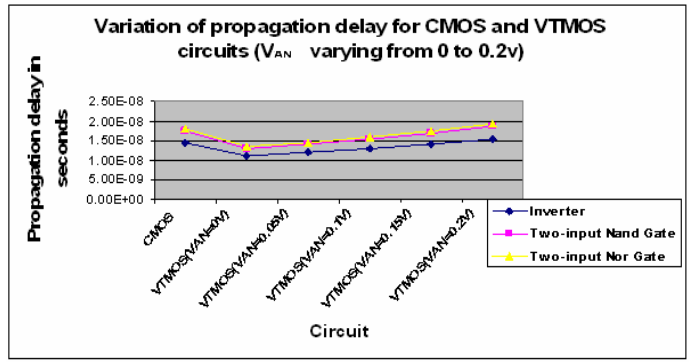

Fig 7(a) Variation of propagation delay for CMOS and VTMOS Circuits ($V_{AN}$ varying from 0 to 0.2V)

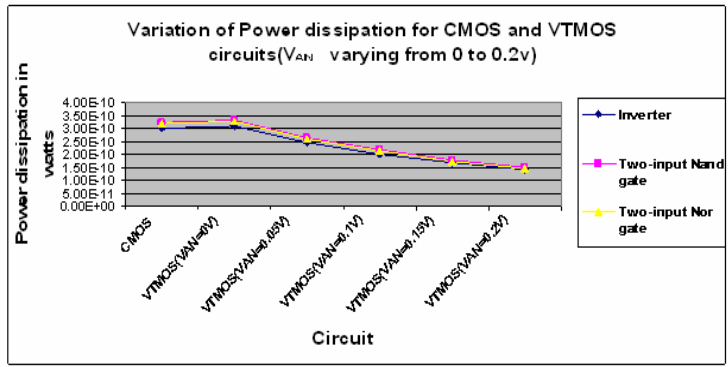

Fig 7(b) Variation of power dissipation for CMOS and VTMOS Circuits ($V_{AN}$ varying from 0 to 0.2V)

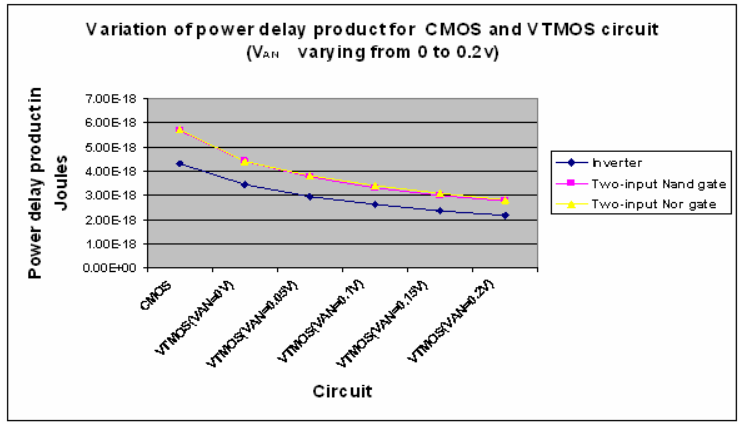

Fig 7(c) Variation of power delay product for CMOS and VTMOS Circuits ($V_{AN}$ varying from 0 to 0.2v)

### 4.2 EFFECT OF FREQUENCY ON THE PERFORMANCE CHARACTERISTICS OF VTMOS CIRCUITS

In order to get an idea of the effect of frequency on the performance characteristics, they are measured at different frequencies ranging from 100 Khz to 8 Mhz. It has been found that the general nature of variation of power dissipation and propagation delays are maintained as those reported at 100 Khz. However, the power dissipation increases with frequency and the variation of power dissipation with frequency for CMOS circuits and VTMOS circuits(0 to 0.2V) is analyzed .The variation of power dissipation with frequency for CMOS and VTMOS two-input NAND is plotted in Fig (8).The same trend follows for VTMOS Inverter and VTMOS two input NOR gates .Propagation delay and logic levels remain almost constant with frequency .At frequencies lower than 8 Mhz, even though power consumption increases with frequency, the gain in the static power dissipation, still provides an advantage compared to CMOS circuits [20,21]. How ever beyond 8 Mhz, the dynamic power is higher than the static power dissipation and there appears to be no advantage of VTMOS compared to CMOS circuits. Therefore it may be noted that for a given technology (feature size) and for a specific W/L of the transistors one can obtain power saving up to certain frequency of operation compared to CMOS. In the particular case discussed in this article, it is found that VTMOS operation in sub threshold region is advantageous up to 8 Mhz operation, beyond this frequency there seems to be no advantage of VTMOS circuits.

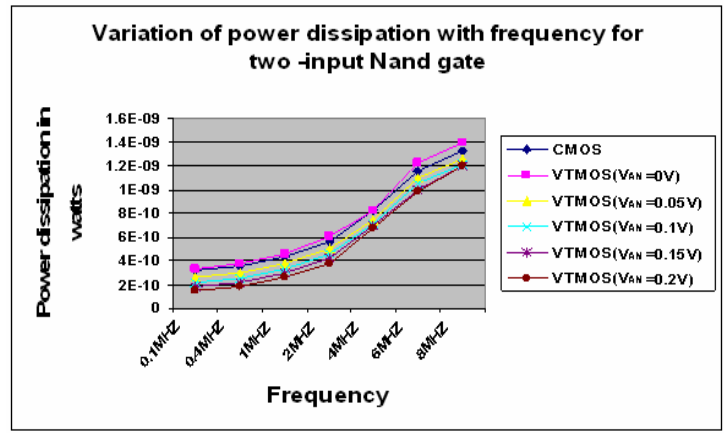

Fig 8 : Variation of power dissipation of Two-input NAND gate with frequency

### 5.0 PERFORMANCE ANALYSIS OF TWO INPUT NAND AND TWO-INPUT NOR LOGIC GATES WITH RANDOM INPUT VECTORS

To evaluate the performance under normal conditions, the performance has been evaluated with random vectors. In this case , each input of NAND and NOR gate is presented with pseudo random inputs and variation of propagation delay, power dissipation ,power delay product and rise and fall time delays are analyzed at 100Khz..The results indicate that with random inputs ,the variation of propagation delay, power dissipation ,power delay product and rise and fall time delays are almost same as in the case with normal inputs .The experiment have been repeated with different random inputs and it is observed that the readings are almost constant .

### 6.0 CONCLUSIONS

This paper reports a modified DTMOS approach which is called VTMOS approach. In these circuits the substrate is operated with a fixed bias ($V_{AN}$ /$V_{AP}$) which results in further reduction in the operating currents compared to DTMOS circuits. The Proposed scheme shows improved power efficiency compared to CMOS and DTMOS circuits , up to a frequency of 8 M h z (for the specific devices used in this Investigation).Using these concepts one may be able to build low power digital circuits like[22,23] which consume lower power than the conventional CMOS and DTMOS circuits.

**REFERENCES**:

[1]    H.Soeleman , K.Roy, and B.C.Paul ( 2001): "Robust Sub threshold Logic for Ultra Low PowerOperation" ,   IEEE trans VLSI system, Vol. 9, PP. 90-99.

[2]    Jabulani:  Nyathi and Brent Bero(2006): "Logic Circuits Operating in Sub Threshold Voltages" , ISLPED  06,  October 4-6,  Tegernsee, Germany.

[3]    C.Hyung, - 11 Kim, H.Soeleman, and K.Roy(2003) : "Ultra-Low Power DLMS Adaptive Filter for Hearing  Aid Applications" ,IEEE transactions on very large scale integration (VLS1) system, vol. 11, No.6,   PP 1058-1067.

[4]    H. Soeleman and K. Roy,  (1999)"Ultra-low power digital subthreshold logic circuits", in Int. Symp.  Low   Power  Electron. Design,  pp. 94–96.

[5]    Fariborz Assaderaghi, Stephen Parke, Dennis Simitsky, Jeffrey Bokor, Ping K. KO. Chenming Hu (1994):"A  Dynamic Threshold Voltage MOSFET (DTMOS) for Very Low Voltage Opertaion" ,IEEE Electron  device letters, vol. 13, No.12.

[6]    Tae-Hyoung Kim, Hanyang Eom, John. Keane and Chris Kim(2006): "Utilizing Reverse Short Channel  Effect   for Optimal Sub Threshold Circuit Design" ,ISLPED 06, October 4-6, 2006. Tegern see,  Germany.

[7]    F. Assaderaghi, D. Sinitsky, S. Parke, J. Bokor, P. K. Ko, and C. Hu:(1994) "A dynamic threshold voltage   MOSFET (DTMOS) for ultra-low voltage operation", in IEDM Tech. Dig.,  pp. 809–812.

[8]    I.Chung, Y.Park, and H. Min (1996): "A new SOI inverter for low power applications",   IEEE  Int.  SOI Conf.,  pp. 20–21.

[9]    A.Drake, K. Nowka, R. Brown  (2003) : " Evaluation of Dynamic-Threshold Logic for Low-Power VLSI   Design   in 0.13um PD-SOI" ,VLSI-SOC pp. 263-266.

[10]    M.R. Casu, G. Masera, G. Piccinini, M. Ruo Roch, M. Zamboni  (2000) :"Comparative Analysis of PD-SOI   Active Body-Biasing Circuits," Proc. IEEE International SOI Conference, pp. 94-95.

[11]    S.Soleimani,A.Sammak,B.Forouzandeh (2009) "A Novel Ultra-Low-Energy Bulk Dynamic Threshold   Inverter   Scheme",Proceedings of the International Multi Conference Of Engineers and Computer   Scientists   Vol 1   IMECS , Hong Kong

[12]    Assaderaghi et al,F (1997) .: "Dynamic Threshold - Voltage Mosfet (DTMOS) for Ultra - Low VoltageVLSI" IEEE trans Electron Devices, Vol. 44, PP 414-422.

[13]    H. Soeleman, K. Roy, B. Paul  (2000)  :"Robust ultra-low power sub-threshold DTMOS logic", in Int.  Symp. Low Power Electron. Design ,pp.25 – 30.

[14]    J. Gil, M. Je, J. Lee and H. Shin (1998) :"A High Speed and Low Power SOI Inverter using Active Body-  Bias", in Int. Symp. Low Power Electron.Design ,  pp.59-63.

[15]    T.Kobayashi, and T.Sakurai (1994) : "Self-adjusting Threshold Voltage Scheme (SATS) for Low- Voltage    High - Speed Operation", in Custom Integrated Circuits Conf.,  pp.271-274.

[16]    S. Narendra, J. Tschanz, J. Hofsheier, B. Bloechel, S. Vangal, Y. Hoskote, S. Tang, D. Somasekhar, A. Keshavarzi, V. Erraguntla, G. Dermer, N. Borkar, S. Borkar and V. De (2004), "Ultra-Low Voltage Circuits  and Processor in 180 nm to 90 nm Technologies with Swapped-Body Biasing Technique," IEEE   International Solid-State Circuits Conference, Vol. 1,. pp. 156-518

[17]   K. von Arnim et al (2005) , "Efficiency of Body Biasing in 90-nm CMOS for Low-Power Digital Circuits,"  Solid-StateCircuits, IEEE Journal of,  pp. 1549 – 1556.

[18]   X. Liu, and S. Mourad  (2000) , "Performance of Submicron CMOS Devices and Gates with Substrate Biasing,"Circuits and Systems, IEEE Int. Symp.,pp. 9 – 12.

[19]   J.M Rabaey: "Digital Integrated circuits: A Design Perspective 1st ed Upper Saddle River" New Jersey:  Prentice Hall, 1996.

[20]    H.Soeleman., K.Roy (2000) .:"Digital CMOS Logic Operation in the Sub-threshold Region," in Tenth Great   Lakes Symp.VLSI,pp.107-112.

[21]    Wet Jin and Philip Chan,CH(1997).: "A Comparative Study of S01 Inverter Circuits for Low Voltage and   Low Power Applications" , Proceedings 1997 IEEE International SOI conference.

[22]    Jaydeep P.Kulkarni,Keejong Kim,and Kaushik Roy (2007).:"A 160 mV Robust Schmitt Trigger Based    Subthreshold SRAM",IEEE Journal Of Solid-State Circuits,VOL 42,NO.10.

[23]     V.Moalemi,Afjali A.Koosha (2007) :"Subthreshold 1-Bit Full Adder Cells in sub-100nm Technologies",Proc.of  the IEEE Computer Society Annual Symposium on VLSI,, pp.514-515.

**BIO-DATA OF AUTHORS**

K.Ragini-Received B.Tech degree from Sri Venkateswara University, Tirupathi in 1990  and  M.Tech degree in Electronics and Communication Engineering from JNTU, Hyderabad .Worked as Assistant Professor in ECE Department  at Chaitanya Bharathi Institute of Technology and Science,Gandipet,Hyderabad from 2000-2005.Then joined as Associate Professor in ECE Department at G.Narayanamma Institute of Technology and Science and presently working in the same institute.Now pursueing Ph.D Under the guidance of DR.M.Satyam, Professor,IIIT,Hyderabad in Low Power VLSI Design.

DR.M.Satyam-Received B.E.(Electronics) from Madras University in 1958.Obtained M.E. and  Ph.D  from Indian Institute of Sciences in 1960 and 1963 respectively.He worked in various capacities at IISc,Bangalore till his retirement. Presently he is working as visiting professor ,IIIT, Hyderabad. He guided 33 Ph.D students and 11 M.S. students. He has 139 publications in various National and International Journals. He has 10 patents to his credit. He is actively engaged in Research and Teaching.

DR.B.C.Jinaga- Born on June, 1950, graduation from Karnataka University, Dharwad, Post Graduation from Regional Engineering, Warangal and Ph.D. from Indian Institute of Technology, Delhi. He has been with JNT University since 21 years. Prof.B.C. Jinaga was occupied various key positions at JNT University . He guided several students for Ph.D. He has published quite number of papers in reputed international and National journals. His research interest includes signal Processing and Coding Techniques. He received Best Teacher Award for the year 2002 - Awarded by Government of Andhra Pradesh.